\documentclass[final,5p,times,twocolumn]{elsarticle}
\usepackage[english]{babel} \usepackage{amsmath} \usepackage{color}
\usepackage{epsfig} \usepackage{graphicx} \usepackage{bm,bbm}
\usepackage{mathtools}

\usepackage{times,amsmath,amssymb,amsfonts}

\journal{Physica A}

\allowdisplaybreaks

\newcommand{\D}{\mathrm{d}}

\definecolor{myred}{RGB}{168,5,14}
\definecolor{myblue}{RGB}{13,13,255}
\definecolor{mygreen}{RGB}{20,150,20}

\newcommand{\EE}{\mbox{$\mathsf E$}} 
\newcommand{\prob}{\mathrm{Prob}}

\begin{document}
\begin{frontmatter}
  
  \title{Phase transitions and gaps in quantum random energy models}
  
  \author[sapienza,infn]{Carlo Presilla}
  \author[ufba]{Massimo Ostilli}
  
  \address[sapienza]{Dipartimento di Fisica, Sapienza Universit\`a di
    Roma, Piazzale A. Moro 2, Roma 00185, Italy}
  \address[infn]{Istituto Nazionale di Fisica Nucleare, Sezione di
    Roma 1, Roma 00185, Italy}
  \address[ufba]{Instituto de
    F\'isica, Universidade Federal da Bahia, Salvador,
    Brazil}

  \begin{abstract}
    By using a previously established exact characterization of the
    ground state of random potential systems in the thermodynamic
    limit, we determine the ground and first excited energy levels of
    quantum random energy models, discrete and continuous. We
    rigorously establish the existence of a universal first order
    quantum phase transition, obeyed by both the ground and the first
    excited states.  The presence of an exponentially vanishing
    minimal gap at the transition is general but, quite interestingly,
    the gap averaged over the realizations of the random potential is
    finite. This fact leaves still open the chance for some effective
    quantum annealing algorithm, not necessarily based on a quantum
    adiabatic scheme.
  \end{abstract}

\end{frontmatter}

\section{Introduction}
The perspective to realize a physical device representing a quantum
computer (QC) has motivated a fervent research activity concerning the
algorithms that could best exploit the intrinsic quantum properties of
such a machine as opposed to the classical ones of current computers.
In particular, there has been a growing interest toward the
possibility to use quantum annealing (QAn)
\cite{Apolloni:1989,Finnila:1994,Kadowaki:1998} as an alternative to
simulated thermal annealing~\cite{Kirkpatrick:1983}.  A pictorial
viewpoint in fact suggests that in order to get the ground state (GS)
of a given classical Hamiltonian $\bm{V}$, the thermal fluctuations,
introduced to avoid the system to be trapped in local minima, could be
replaced by quantum fluctuations able to outperform the former due to
tunneling effects.  Usually, QAn is associated with quantum adiabatic
(QAd) algorithms~\cite{Farhi:2000,Farhi:2001,Santoro:2006}.  The idea
is to implement an interpolating Hamiltonian
$\bm{H}(\Gamma)=\bm{V}+\Gamma \bm{K}$, where $\bm{K}$ is an operator
which does not commute with $\bm{V}$.  The adiabatic theorem ensures
that for sufficiently slow changes of the parameter $\Gamma$ the
interpolating system remains in its GS so that the original GS of
$\bm{V}$ can be recovered in the limit $\Gamma\to 0$.  However, for
many interesting problems $\bm{V}$, the interpolating Hamiltonian
$\bm{H}(\Gamma)$ is likely to undergo a first order quantum phase
transition at some value $\Gamma=\Gamma_\mathrm{c}$, where the energy
gap $\Delta$ between the first excited state (FES) and the GS becomes
exponentially small in the system size $N$
\cite{Jorg:2008,Jorg:2010,Warzel2015}.  In this case, a QAd decrease
of $\Gamma$ starting from some value $\Gamma>\Gamma_\mathrm{c}$, where
the GS of $\bm{H}(\Gamma)$ is found with ease, requires an
exponentially long time.  Otherwise, the system evolves into a
combination of several instantaneous eigenstates of $\bm{H}(\Gamma)$
and when $\Gamma\to 0$ there is a finite probability to attain a state
of $\bm{V}$ different from its GS. For $\bm{V}$ with a glassy energy
landscape, ``quantum is better'' may be untrue~\cite{Santoro:2005-2}.

The aim is this paper is threefold, we show that: (i) this phase
transition scenario, characterized by a normal (paramagnetic) phase,
for $\Gamma>\Gamma_\mathrm{c}$, and a condensed phase, for
$\Gamma<\Gamma_\mathrm{c}$, is in fact universal, being valid for any
hopping operator $\bm{K}$ and any potential operator $\bm{V}$; (ii) in
the case of random potentials, where one must perform a quenched
average over the different realizations of $\bm{V}$, or, instances,
the average of the GS energy determines $\Gamma_c$ (see
Eq.~\ref{critical}), the average of the gap remains finite and, with
respect to $\Gamma$, is constant in the condensed phase, and linear in
the normal phase; (iii) for any instance there exists a value
$\Gamma_\mathrm{min}$ where the gap is exponentially small with the
system size $N$, and $\Gamma_\mathrm{min}\to\Gamma_\mathrm{c}$ for
$N\to\infty$.  All the theoretical analysis is supported by unbiased
numerical simulations.  In the conclusions we shortly discuss the
implications of points (i)-(iii) which make impossible to realize an
efficient quantum adiabatic algorithm, but still leave open the chance
for some effective more complicated quantum annealing
scheme~\cite{Farhi:2009}.

We analyze the case in which $\bm{V}$ is a generic random potential
with a discrete or continuous distribution of the levels.  For any
choice of $\bm{K}$, provided that $\bm{K}$ has zero diagonal elements
in the representation in which $\bm{V}$ is diagonal,
$\bm{H}(\Gamma)=\bm{V}+\Gamma \bm{K}$ is a quantum random energy model
(QREM) belonging to the class of systems studied in~\cite{OP:2006}.
The term QREM refers to the quantum counterpart of the classical
random energy model (REM), the model defined without the ``hopping''
operator $\bm{K}$, namely, $\bm{H}=\bm{V}$. In its original version,
the REM was proposed with $\bm{V}$ having Gaussian distributed levels
and represents a well known toy model for spin glasses
\cite{Derrida:1980}. The corresponding QREM has been first studied
perturbatively and numerically in~\cite{Jorg:2008}.

In Ref.~\cite{OP:2006} we have exactly characterized the GS and found
a sufficient condition for the existence of a first order quantum
phase transition for systems in which $\bm{V}$ is an arbitrary random
potential.  Here, we extend these results to study in detail both the
GS and the FES energies of a generic QREM.

\section{Ground state of random potential systems}
In~\cite{OP:2006} we have determined the exact GS of a class of
Hamiltonian models defined by an arbitrary hopping operator $\bm{K}$,
i.e., an off diagonal matrix of dimension $M$, and a random potential
$\bm{V}$, i.e., a diagonal matrix with $M$ i.i.d. random values $V$
extracted according to an arbitrary probability distribution $P(V)$.
As usual, we will denote the system corresponding to a particular
realization of the $V$ values as an \textit{instance} and
$\mathsf{E}(\cdot) = \int \cdot\ P(V) \D{V}$ will stand for the
expectation over all possible instances.  For the mentioned class of
models, \emph{in the thermodynamic limit} the energy $E_0$ of the GS
is related to the lowest level $E_0^{(0)}$ of the hopping operator by
\begin{align}
  \int \frac{P(V)}{E_0-V}~\D{V} = \frac{1}{E_0^{(0)}}, \qquad E_0 \leq
  \mathsf{E}(V_{(1)}),
  \label{maineq}
\end{align}
where $\mathsf{E}(V_{(k)})$ is the expectation of the $k$-th order
statistic associated with the distribution $P(V)$, i.e., the
expectation of the $k$-th smallest value among the $M$ values of $V$
drawn according to $P(V)$~\cite{JKK:2005}.  Note that, whereas
$E_0^{(0)}$ is deterministic, $E_0$ is stochastic and
Eq.~(\ref{maineq}) is actually an equation for the expectation
$\textsf{E}(E_0)$.  However, we assume that $E_0$ is self-averaging
and this justifies the above notation.

The derivation of Eq.~(\ref{maineq}) follows from an exact
probabilistic representation of the time evolution of a quantum system
in terms of a proper collection of independent Poisson
processes~\cite{DeAJLS:1983,DeAJL:1998,BPDeAJL:1999}. In fact, by
using this representation, the lowest eigenvalue of
$\bm{H}=\bm{V}+\Gamma \bm{K}$ can be expressed as the solution of a
scalar equation written in terms of the asymptotic probability density
of the potential and hopping frequencies (frequencies of the values of
$\bm{V}$ and $\bm{K}$ realized in the probabilistic representation of
an infinitely long time evolution)~\cite{OP:2006}. It happens that,
for the above mentioned class of systems with a random potential, as
well as for the uniformly fully connected models, in the thermodynamic
limit the asymptotic probability density of the potential and hopping
frequencies is exactly given by a multinomial. The equation for the
ground-state level of $\bm{H}$ then greatly simplifies and, in the
thermodynamic limit, takes on the form of
Eq.~(\ref{maineq})~\cite{OP:2006}. We assume a non-trivial
thermodynamic limit, namely, a limit in which the lowest levels of
$\bm{H}$, $\bm{V}$ and $\bm{K}$ diverge with the same speed (the most
interesting case being the one in which such levels are all extensive
in $N$).  Apart from this, the distribution $P(V)$ remains completely
arbitrary.

Equation~(\ref{maineq}) has a simple interpretation: the average of
the inverse hopping energy, estimated as $1/(E_0-V)$, must coincide
with the inverse of the actual hopping energy, namely, $1/E_0^{(0)}$,
with the constraint that $E_0$ does not exceed the averaged minimum
potential.  It may happen that the solution of the integral problem in
Eq.~(\ref{maineq}) becomes incompatible with the constraint $E_0 \leq
\mathsf{E}(V_{(1)})$. This may occur, for instance, when some
parameter of the model is changed below a threshold, e.g., $\Gamma <
\Gamma_\mathrm{c}$ for the Hamiltonian $\bm{H}(\Gamma)=\Gamma
\bm{K}+\bm{V}$.  Then, the GS of an instance condensates, for $\Gamma
< \Gamma_\mathrm{c}$, into the eigenstate of $\bm{V}$ corresponding to
the minimum value $V$ realized in that instance.  As a result,
$E_0(\Gamma)=\mathsf{E}(V_{(1)})$ for any $\Gamma <\Gamma_\mathrm{c}$
and we obtain a condensation in the space of
states~\cite{OP:condensation}. Note that each instance condensates
into a different state, however, this state corresponds, for all
instances, to that of minimum potential. A sufficient condition for
this first order quantum phase transition to take place is that
$P(\mathsf{E}(V_{(1)}))\to 0$ in the thermodynamic limit
\cite{OP:2006}.

When the above results are applied to a quantum mechanical model, $M$
stands for the dimension of the Fock space and, in general, we have $M
\gg N$, where $N$ is the size (number of particles) of the system.
The phase transition is determined by the condition
$E_0=\mathsf{E}(V_{(1)})$, where $E_0$ is the solution of the integral
problem in Eq.~(\ref{maineq}), namely
\begin{align}
  \int \frac{P(V)}{\mathsf{E}(V_{(1)})-V}~\D{V} = \frac{1}{E_0^{(0)}}.
  \label{critical}
\end{align}
This is the equation which allows us to find the critical point of the
phase transition in terms of the ``hopping'' parameter $\Gamma$ or,
possibly, in terms of other parameters contained in the probability
distribution $P(V)$.

\section{Discrete and continuous QREMs}
Let us consider the Hamiltonian $\bm{H}=\Gamma \bm{K}+\bm{V}$ acting
on the $M=2^N$ spin states
\begin{align}
  | \bm{n} \rangle = | n_1,n_2,\dots,n_N \rangle = | n_1\rangle
  \otimes |n_2\rangle \otimes \dots \otimes |n_N \rangle,
\end{align}
where $|n_i\rangle$, with $n_i=\pm 1$, is the eigenstate of the Pauli
matrix $\sigma^z_i$ acting on the $i$-th spin and $i=1,\dots,N$,
$\sigma^z_i |n_i\rangle = n_i |n_i\rangle$.  The hopping operator
$\bm{K}$ is chosen as the sum of single-flip operators
\begin{align*}
  \bm{K} = - \sum_{i=1}^N \mathbbm{1}_1 \otimes \ldots \otimes
  \mathbbm{1}_{i-1} \otimes \sigma^x_i \otimes \mathbbm{1}_{i+1} \dots
  \otimes \ldots \mathbbm{1}_N,
\end{align*}
$\sigma^x_i$ being the $x$ Pauli matrix acting on the $i$-th spin.
More general forms for $\bm{K}$ can be considered and tackled in a
similar way.  The potential $\bm{V}$ is a diagonal operator with
elements $V_{\bm{n}}=\langle \bm{n}|\bm{V}| \bm{n} \rangle$ which are
i.i.d. random variables drawn with distribution $P(V)$.  As an example
of discrete QREMs we will consider the case in which $V$ takes the
values $V_k=-J(N-2k)$, $k=0,1,\ldots,N$, with probability mass
function $P(V_k) = 2^{-N} \binom{N}{k}$.  We will refer to this model
as the binomial QREM.  As an example of continuous models we will
assume $V\in (-\infty,+\infty)$ with probability density function
$P(V) = (\pi N J^2)^{-1/2} \exp(-V^2/(NJ^2))$.  We will refer to this
model as the Gaussian QREM.

  In the thermodynamic limit $N\to\infty$, the distribution $P(V)$
  concentrates around some value $V_\mathrm{sp}$ where
  $P'(V_\mathrm{sp})=0$ and $P''(V_\mathrm{sp})<0$. The integral in
  Eq.~(\ref{maineq}) can be then performed, exactly, by the
  saddle-point method, namely,
  \begin{align}
    \int \frac{P(V)}{E_0-V}\D{V} = \frac{1}{E_0-V_\mathrm{sp}}.
  \end{align}
  For symmetric distributions, $P(-V)=P(V)$, we have $V_\mathrm{sp}=0$
  and Eq.~(\ref{maineq}) amounts to $E_0 = \min(
  E_0^{(0)},\mathsf{E}(V_{(1)}))$.  

The spectrum of $\Gamma \bm{K}$ is trivial and, in particular, the GS
level is $E_0^{(0)} = - \Gamma N$.  Equation~(\ref{maineq}) thus
provides
\begin{align}
  E_0 = \left\{
    \begin{array}{ll}
      \mathsf{E}(V_{(1)}) \qquad& \Gamma < \Gamma_\mathrm{c}
      \\
      -\Gamma N \qquad& \Gamma \geq \Gamma_\mathrm{c}
    \end{array}
  \right. ,
  \label{E0}
\end{align}
where, according to Eq.~(\ref{critical}),
$\Gamma_\mathrm{c}=\lim_{N\to\infty}-\mathsf{E}(V_{(1)})/N$.  Stated
in this form, the result applies to any QREM.  In the binomial model
$\mathsf{E}(V_{(1)}) = -JN+O(1)$, whereas in the Gaussian model
$\mathsf{E}(V_{(1)}) = -JN\sqrt{\ln 2}+O(1)$.  The critical hopping
parameter separating the condensed and the normal phases is then
$\Gamma_\mathrm{c}=J$ and $\Gamma_\mathrm{c}=J\sqrt{\ln 2}$,
respectively.

By introducing auxiliary Hamiltonians, we can use again
Eq.~(\ref{maineq}) to evaluate the eigenvalues of $\bm{H}$
corresponding to excited states.  Suppose, to begin with, that we are
in the condensed phase $\Gamma N < -\mathsf{E}(V_{(1)})$.  For a given
realization of the random potential $\bm{V}$, the GS of $\bm{H}$ is
$|\bm{n}_1\rangle$, where $\bm{n}_1$ is the spin configuration
associated with the smallest value of the potential, i.e.,
$V_{\bm{n}_1} =\min_{\bm{n}} V_{\bm{n}}$.  For simplicity, we
  suppose that $V_{\bm{n}_1}$ is non-degenerate, a condition which is
  almost surely met in the Gaussian QREM. For $V_{\bm{n}_1}$
  degenerate a similar argument applies. We then introduce the
Hamiltonian
\begin{align}
  \tilde{\bm{H}} = \bm{H} - V_{\bm{n}_1} |\bm{n}_1
  \rangle\langle\bm{n}_1|,
\end{align}
whose lowest eigenvalue $\tilde{E}_0$ coincides with the level $E_1$
of $\bm{H}$.  Note that $\tilde{\bm{H}}$ describes a system with
random potential $\tilde{V}$ having distribution
$\tilde{P}(\tilde{V})$ such that $\tilde{\mathsf{E}}(\tilde{V}_{(1)})
= \mathsf{E}(V_{(2)})$.  Moreover, the operators $\tilde{\bm{H}}$ and
$\bm{H}$ have the same non-diagonal part, i.e.,
$\tilde{E}_0^{(0)}=E_0^{(0)}$.  From Eq.~(\ref{maineq}) applied to the
Hamiltonian $\tilde{\bm{H}}$ we thus find
\begin{align}
  \tilde{E}_0 = \left\{
    \begin{array}{ll}
      \mathsf{E}(V_{(2)}) \qquad& \Gamma < -\mathsf{E}(V_{(2)})/N
      \\      
      -\Gamma N \qquad& \Gamma \geq -\mathsf{E}(V_{(2)})/N
    \end{array}
  \right. .
  \label{E1frozen}
\end{align}

We proceed with a similar analysis in the normal phase.  For $\Gamma N
\gg -\mathsf{E}(V_{(1)})$, the GS of $\bm{H}$ approaches the GS of
$\bm{K}$, namely
$|0_{\bm{K}}\rangle=2^{-N/2}\sum_{\bm{n}}|\bm{n}\rangle$.  Observing
that $\mathsf{E}(\langle 0_{\bm{K}} |\bm{H}| 0_{\bm{K}} \rangle)
=-\Gamma N+\mathsf{E}(V)=-\Gamma N$ exactly for any $\Gamma$, we
assume $|0_{\bm{K}}\rangle$ to represent an effective GS in all the
normal region. We then introduce the Hamiltonian
\begin{align}
  \tilde{\bm{H}} = \bm{H} + \Gamma N |0_{\bm{K}} \rangle\langle
  0_{\bm{K}}|,
\end{align}
which, we guess, in all the normal region, has the average lowest
eigenvalue $\tilde{E}_0$ coincident with the average level $E_1$ of
$\bm{H}$.  We split $\tilde{\bm{H}}$ as
\begin{align}
  \tilde{\bm{H}} = \tilde{\bm{V}} + \Gamma \tilde{\bm{K}},
\end{align}
where $\tilde{\bm{V}}=\bm{V}+\Gamma N 2^{-N}$ is a random potential
with distribution $\tilde{P}(\tilde{V})=P(V)$, and
$\tilde{\bm{K}}=\bm{K} + N 2^{-N} (\bm{W}-\bm{1})$ a hopping operator
with $\bm{W}$ defined by matrix elements $W_{\bm{n},\bm{n}'}=1$.  Let
$|1_{\bm{K}}^{(i)}\rangle=\sigma_i^z |0_{\bm{K}}\rangle$,
$i=1,\dots,N$, be the $N$-degenerate FESs of $\bm{K}$, namely,
$\bm{K}|1_{\bm{K}}^{(i)}\rangle=-\Gamma
(N-2)|1_{\bm{K}}^{(i)}\rangle$.  Similarly, let
$|2_{\bm{K}}^{(i,j)}\rangle=\sigma_i^z\sigma_j^z |0_{\bm{K}}\rangle$,
$i,j=1,\dots,N$, with $i\neq j$, be the $N(N-1)/2$-degenerate second
exited states of $\bm{K}$, namely,
$\bm{K}|2_{\bm{K}}^{(i,j)}\rangle=-\Gamma
(N-4)|2_{\bm{K}}^{(i,j)}\rangle$, and so on.  By using
$\bm{W}|0_{\bm{K}}\rangle=2^N|0_{\bm{K}}\rangle$ and
$\bm{W}|1_{\bm{K}}^{(i)}\rangle=\bm{W}|2_{\bm{K}}^{(i,j)}\rangle=\ldots=0$,
we find that any $|1_{\bm{K}}^{(i)}\rangle$ is a GS of
$\tilde{\bm{K}}$ with eigenvalue $-(N-2)-N2^{-N}$.  By inserting
$\tilde{E}_0^{(0)}=-\Gamma (N-2+N 2^{-N})$, and
$\tilde{\mathsf{E}}(\tilde{V}_{(1)}) = \mathsf{E}(V_{(1)}) + \Gamma N
2^{-N}$ into Eq.~(\ref{maineq}) for $\tilde{\bm{H}}$, we get
\begin{align}
  \tilde{E}_0 = \left\{
    \begin{array}{ll}
      \mathsf{E}(V_{(1)})+\Gamma N2^{-N} \qquad& 
      \Gamma  < \frac{-\mathsf{E}(V_{(1)})}{N-2+N2^{-N}}
      \\
      -\Gamma (N-2) \qquad& 
      \Gamma \geq \frac{-\mathsf{E}(V_{(1)})}{N-2+N2^{-N}}
    \end{array}
  \right. .
  \label{E1para}
\end{align}

\begin{figure}
  \includegraphics[width=1\columnwidth,clip]{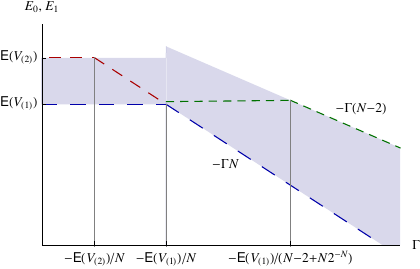}
  \caption{Lowest two energy levels $E_0$ and $E_1$ (bottom and top
    edges of the shaded area, respectively) for a QREM in the
    thermodynamic limit as a function of the hopping parameter
    $\Gamma$. The blue long-dashed line is the value of $E_0$ given by
    Eq.~(\ref{E0}), whereas the red medium-dashed and green
    short-dashed lines represent the finite-$N$ estimate for $E_1$
    given by Eqs.~(\ref{E1frozen}) and~(\ref{E1para}), respectively.
    As $N\to\infty$ the critical points (corresponding to the dotted
    vertical lines) concur at
    $\Gamma_\mathrm{c}=-\mathsf{E}(V_{(1)})/N$ where $E_1$ develops a
    $O(1)$ discontinuity.}
  \label{QREM.common.features}
\end{figure}

Finally, we combine Eqs.~(\ref{E1frozen}) and~(\ref{E1para}) to obtain
$E_1$, see Fig.~\ref{QREM.common.features}.  Observing that
$\lim_{N\to\infty}\left(\mathsf{E}(V_{(2)})-\mathsf{E}(V_{(1)})\right)/N=0$,
we conclude that in the thermodynamic limit
\begin{align}
  E_1 = \left\{
    \begin{array}{lr}
      \mathsf{E}(V_{(2)}) \qquad& \Gamma < \Gamma_\mathrm{c}
      \\
      -\Gamma (N-2) \qquad& \Gamma \geq \Gamma_\mathrm{c}
    \end{array}
  \right. .
  \label{E1}
\end{align}

Equation~(\ref{maineq}) and hence $E_0$ and $E_1$ found above are
exact up to terms $O(1)$~\cite{OP:2006}.  However, the $O(1)$ errors
obtained for $E_0$ and $E_1$ match when $N\to\infty$.  It follows that
for $N\to\infty$ the average gap $\mathsf{E}(\Delta)=E_1-E_0$ is
\begin{align}
  \mathsf{E}(\Delta) = \left\{
    \begin{array}{ll}
      \mathsf{E}(V_{(2)})-\mathsf{E}(V_{(1)}) \qquad& \Gamma < \Gamma_\mathrm{c}
      \\
      2\Gamma \qquad& \Gamma \geq \Gamma_\mathrm{c}
    \end{array}
  \right. .
  \label{gap}
\end{align}
In the condensed phase, the average gap is a constant which amounts to
$\mathsf{E}(\Delta)/J=2/e$ in the binomial QREM and
$\mathsf{E}(\Delta)/J \simeq 0.62$ in the Gaussian QREM.  These
results immediately follow from the evaluation of the first two order
statistics associated with the chosen $P(V)$.  In the case of the
binomial distribution, we have
\begin{align}
  \EE(V_{(1)}) &= -JN \prob(V_{(1)}=-JN) \nonumber \\ &\quad -J(N-2)
  \prob(V_{(1)}=-J(N-2)) +\dots \nonumber \\ &= -JN + J\frac{2}{e},
\end{align}
\begin{align}
  \EE(V_{(2)}) &= -JN \prob(V_{(2)}=-JN) \nonumber \\ &\quad -J(N-2)
  \prob(V_{(2)}=-J(N-2)) +\dots \nonumber \\ &= -JN + J\frac{4}{e},
\end{align}
where the probabilities have been estimated in the limit $N\to\infty$,
namely,
\begin{align}
  \label{p1.1}
  &\prob(V_{(1)}=-JN) = 1- \frac{1}{e},
  \\
  \label{p1.2}
  &\prob(V_{(1)}=-J(N-2)) = \frac{1}{e},
  \\
  \label{p2.1}
  &\prob(V_{(2)}=-JN) = 1- \frac{2}{e},
  \\
  \label{p2.2}
  &\prob(V_{(2)}=-J(N-2)) = \frac{2}{e}.
\end{align}
Note that, in this limit, $\prob(V_{(1)}\geq-J(N-4))$ and
$\prob(V_{(2)}\geq-J(N-4))$ vanish.  In the case of the Gaussian
distribution, the $O(1)$ contributions to $\mathsf{E}(V_{(1)})$ and
$\mathsf{E}(V_{(2)})$ have been evaluated numerically, the difference
being close to $0.62~J$.

\begin{figure}[t]
  \includegraphics[width=1\columnwidth,clip]{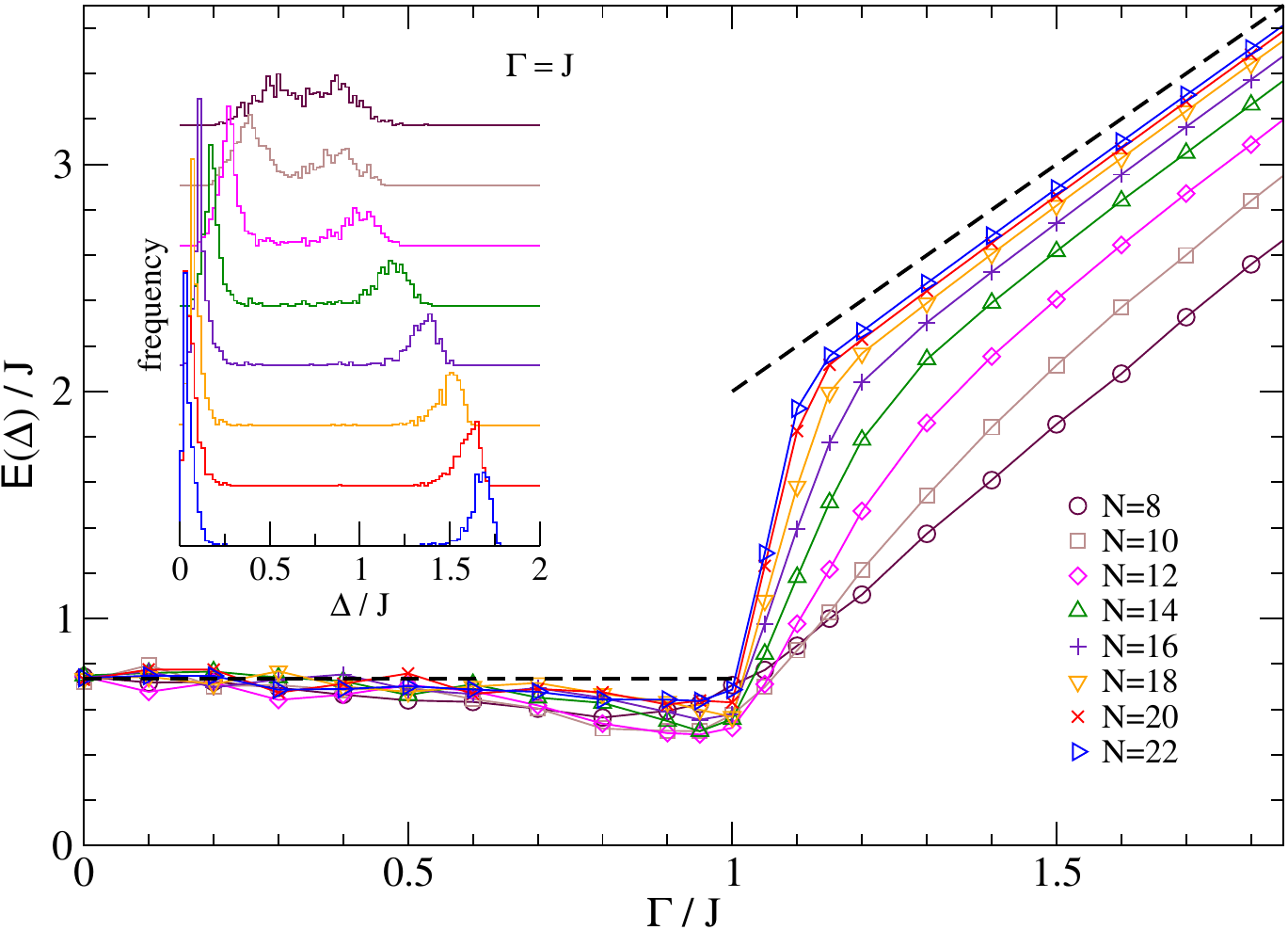}
  \caption{ Average gap $\mathsf{E}(\Delta)$ in the binomial QREM as a
    function of $\Gamma$ for different system sizes $8 \leq N \leq
    22$.  The dashed line is the predicted value in the thermodynamic
    limit whereas for each size $N$ the symbols joined by straight
    lines are the averages from exact numerical diagonalization of
    1000 instances.  Inset: histograms of the frequencies of the gap
    values evaluated at $\Gamma = \Gamma_\mathrm{c}$ (size $8 \leq N
    \leq 22$ from top to bottom, curves shifted for the sake of
    clarity).}
  \label{gb}
\end{figure}

\begin{figure}[t]
  \includegraphics[width=1\columnwidth,clip]{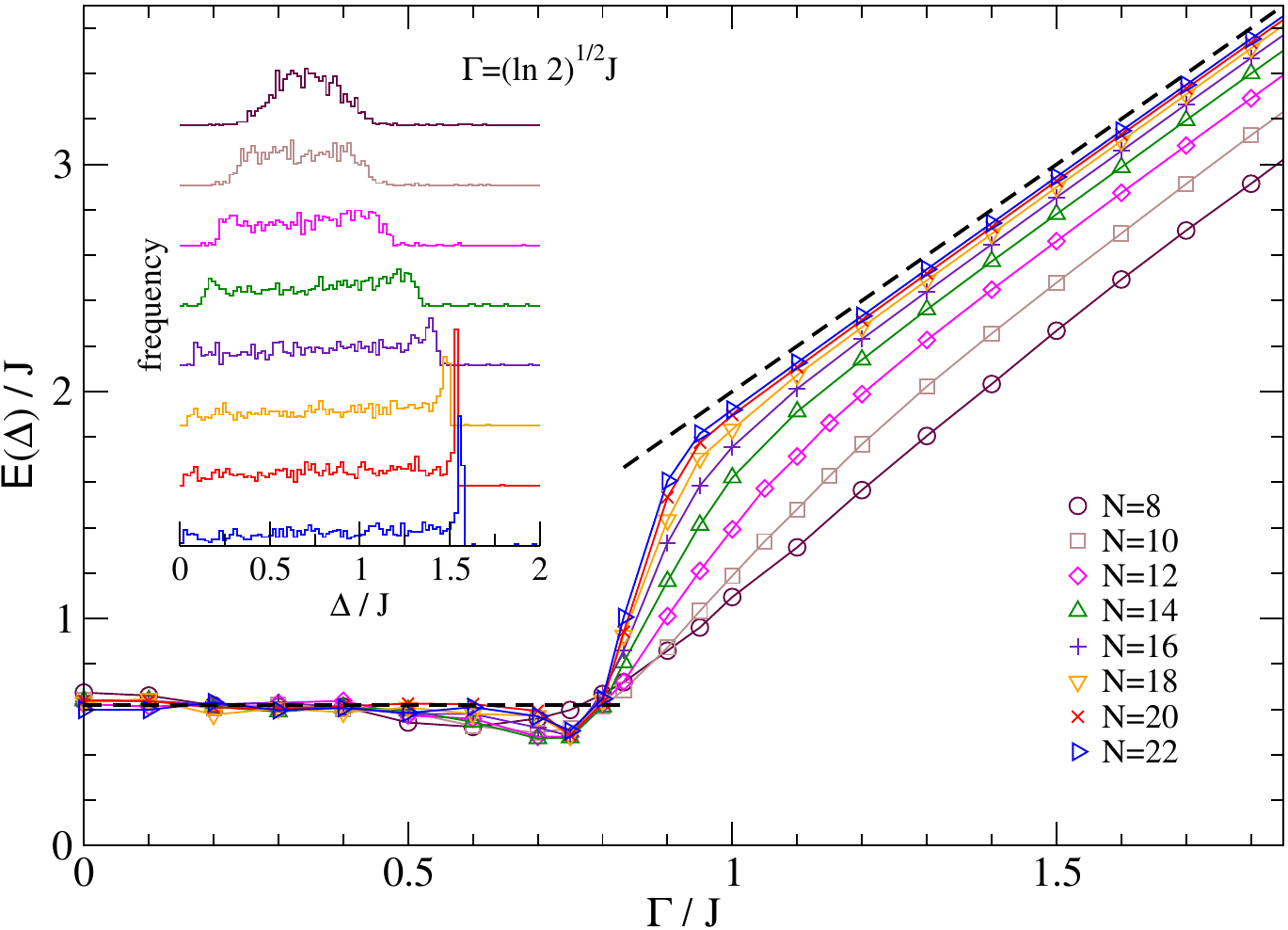}
  \caption{As in Fig.~\ref{gb} for the Gaussian QREM.}
  \label{gg}
\end{figure}

The results of Eqs.~(\ref{E0}),~(\ref{E1}) and~(\ref{gap}) have been
compared with those from exact numerical diagonalizations of $\bm{H}$
for different values of the system size $N$.  In Figs.~\ref{gb} and
\ref{gg} we show the behavior of the average gap in the binomial and
Gaussian QREMs, respectively.  In both cases the data obtained for $N$
from 8 to 22 exhibit a systematic convergence toward the thermodynamic
limit~(\ref{gap}).


Our numerical data confirm that, as usually assumed, the extensive
quantities $E_0$ and $E_1$ are self-averaging.  For the gap, on the
other hand, we observe a different behavior.  In the normal phase,
$\Delta$ is self-averaging whereas this property is lost when the
system condensates.  In the insets of Fig.~\ref{gb} and \ref{gg} we
show the frequencies of the gap values at the critical point
$\Gamma=\Gamma_\mathrm{c}$ for the binomial and Gaussian QREMs,
respectively.  These frequencies, evaluated from a set of 1000
instances, have been checked to be a good approximation of the exact
distributions of $\Delta$ at the chosen sizes $N$.  It is evident that
$\lim_{N\to\infty} \mathrm{var}(\Delta)/ \mathsf{E}(\Delta)^2 \neq 0$,
i.e., $\Delta$ is not self-averaging.  The same conclusion is reached
for both binomial and Gaussian QREMs in the whole range $ 0 \leq
\Gamma \leq \Gamma_\mathrm{c}$.

The gap distributions for the binomial and the Gaussian QREMs have
well distinct shapes in the condensed phase.  In the discrete model,
when $\Gamma$ is decreased below $\Gamma_\mathrm{c}$ the gap values
concentrate around $\Delta=0,2J,4J,\dots$. Including only terms not
vanishing in the limit $N\to\infty$, at $\Gamma=0$ the gap
distribution becomes $ q_N(\Delta)=r_N \delta(\Delta) + (1-r_N)
\delta(\Delta-2J)$, where $r_N$ can be calculated analytically and for
$N$ large approaches $1-1/e$. This behavior is qualitatively not
different from that shown in the inset of Fig.~\ref{gb} for
$\Gamma=\Gamma_\mathrm{c}$.  In the continuous model, the gap values
are evenly distributed in the range $0 \leq \Delta \leq
2\Gamma_\mathrm{c}$ for $\Gamma \lesssim \Gamma_\mathrm{c}$. The
distribution smoothly deforms by decreasing $\Gamma$ and at $\Gamma=0$
is exactly described by the function
\begin{align}
  q_N(\Delta) =\ & \theta(\Delta) \frac{2^N(2^N-1)}{\pi\sqrt{NJ^2}}
  \int_{-\infty}^{+\infty} e^{-t^2-(t+\Delta/\sqrt{NJ^2})^2} \nonumber
  \\ &\times [ \mathrm{Erfc}(t+\Delta/\sqrt{NJ^2})/2 ]^{2^N-2} \D{t}.
  \label{qNg}
\end{align}
The logarithm of~(\ref{qNg}) is well approximated by the first two
terms of its Taylor expansion at $\Delta=0$.  By using the
normalization condition, we then have $q_N(\Delta) \simeq
q_N(0)\exp[-q_N(0)\Delta]$, where $q_N(0)$ quickly saturates for
$N\to\infty$ to a constant value near
$1/\mathsf{E}(\Delta)=1/(0.62~J)$.

\section{Minimal gap}
It is interesting to understand how the flat trend of $\EE(\Delta)$
observed in the whole condensed region, is attained from the actual
behavior of the gap of the single instances. In fact, at least for the
Gaussian QREM, it has been shown by numerical and theoretical
arguments~\cite{Jorg:2008}, as well as rigorously demonstrated
\cite{Warzel2015}, that, close to the critical point, the gap of each
single instance has a rather different behavior: it vanishes
exponentially by increasing the size $N$.
\begin{figure}[t]
  \includegraphics[width=1\columnwidth,clip]{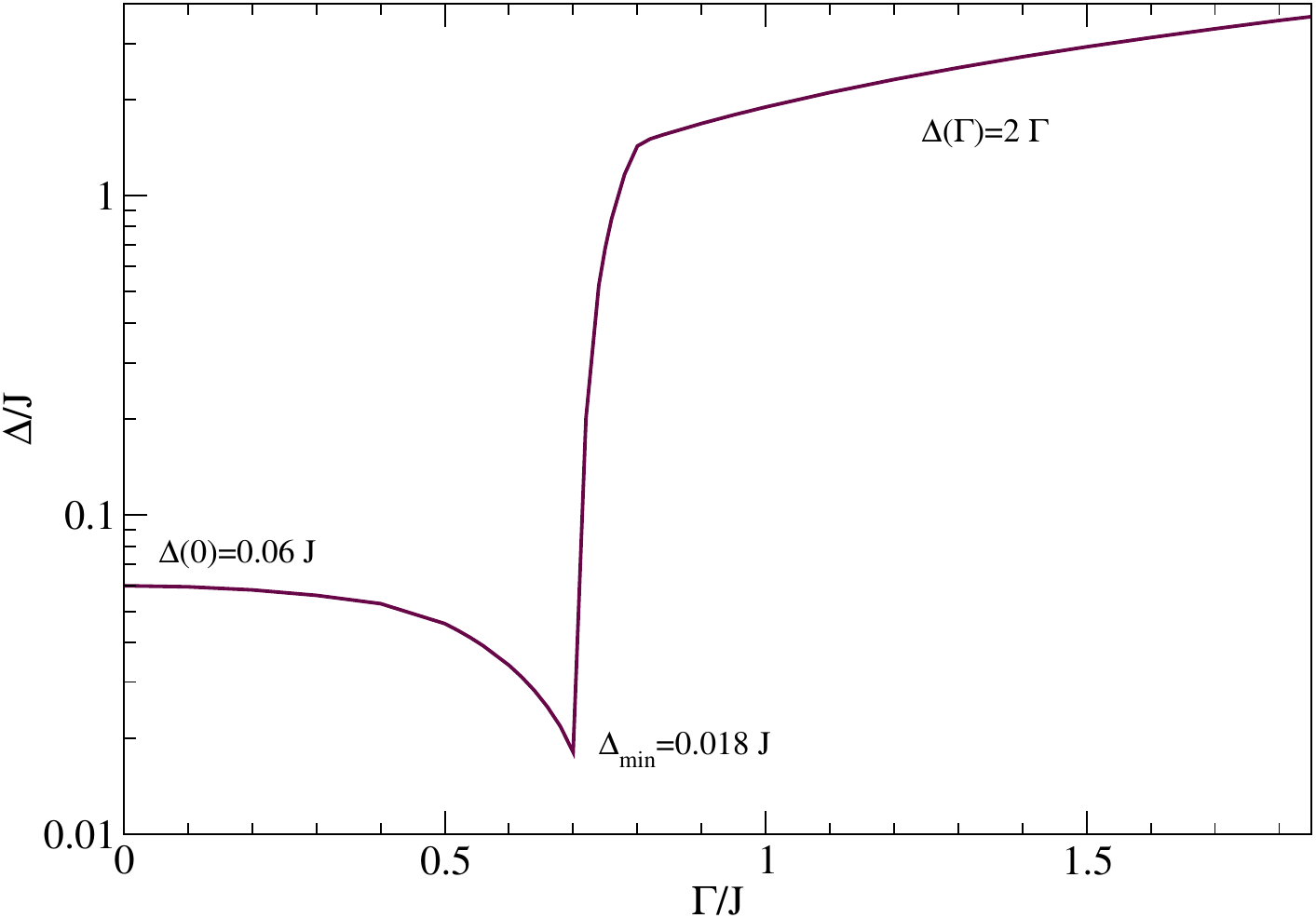}
  \caption{ Gap $\Delta$ of an instance as a function of $\Gamma$ for
    a Gaussian QREM model with $N=20$ spins.  Similar results are
    obtained for a binomial QREM.}
  \label{qa}
\end{figure}
In fact, for any instance of size $N$, by varying $\Gamma$ the gap
$\Delta(\Gamma)$ has a minimum at $\Gamma=\Gamma_\textrm{min}$ as
shown in Fig.~\ref{qa}. Let us define $\Delta_\textrm{min} =
\min_{\Gamma \geq 0}\Delta(\Gamma)$. Theoretical arguments, checked
numerically for both Gaussian and binomial QREMs, show that (a)
$\Delta_\textrm{min}$ and $\Gamma_\textrm{min}$ are self-averaging
quantities, (b) $\mathsf{E}(\Delta_\textrm{min}) \simeq
2\Gamma_\mathrm{c} N2^{-N/2}$ for $N$ large and (c) $\lim_{N\to\infty}
\mathsf{E}(\Gamma_\textrm{min}) = \Gamma_\mathrm{c}$ with
$\mathrm{var}(\Gamma_\textrm{min}) \simeq (\Gamma_\mathrm{c}/N)^2$ for
$N$ large.

In principle, the shape of $\Delta(\Gamma)$ for a given instance can
be accurately reproduced by the variational Ritz method writing the
lowest eigenstates of $\bm{H}(\Gamma)$, $|E_n(\Gamma)\rangle$ with
$n=0,1$, as a superposition of the lowest eigenstates of $\bm{K}$ and
$\bm{V}$.  In the crudest approximation, namely,
\begin{align}
  |E_n(\Gamma)\rangle = a_n(\Gamma) |0_{\bm{K}}\rangle+b_n(\Gamma)
  |\bm{n}_1 \rangle,
\end{align}
by using $\langle 0_{\bm{K}}|V|0_{\bm{K}}\rangle=0$, valid whenever
$P(-V)=P(V)$, one obtains the analytical estimate
\begin{eqnarray}
  \Delta(\Gamma) =
  \sqrt{(N\Gamma+V_{\bm{n}_1})^2-4N\Gamma V_{\bm{n}_1}2^{-N}},
  \label{Delta(Gamma)}
\end{eqnarray}
which gives
\begin{align}
  \Gamma_\mathrm{min} = -V_{\bm{n}_1}/N
\end{align}
and
\begin{align}
  \Delta_\mathrm{min} = -2V_{\bm{n}_1} 2^{-N/2}.
\end{align}
Properties (a), (b) and (c) follow straightaway from these two
expressions observing that $V_{\bm{n}_1}=V_{(1)}$ and, for
$N\to\infty$, $\mathsf{E}(V_{(1)}) = -\Gamma_\mathrm{c}N+O(1)$,
whereas $\mathrm{var}(V_{(1)})$ approaches a constant value close to
$\Gamma_\mathrm{c}^2$.

We are now in a position to discuss how the results of Figs.~\ref{gb}
and \ref{gg} are recovered from the behavior of the gap of the
individual instances.

In the region $\Gamma < \Gamma_\mathrm{c}$, where the gap is not
self-averaging, $\mathsf{E}(\Delta)$ must loose memory of the
instance-dependent minimum of $\Delta$. Close to the critical point,
this is quantitatively understood as follows.  For $N$ large and
$\Gamma \simeq \Gamma_\mathrm{min} = -\EE(V_{(1)})/N$, according to
Eq.~(\ref{Delta(Gamma)}), we have $\Delta \simeq
|-\EE(V_{(1)})+V_{(1)}|$. We have to average this expression over all
values of $V_{(1)}$.  For the Gaussian QREM, neglecting large
deviatons, the probability distribution of $V_{(1)}$ is well
approximated by a Gaussian centered at $\EE(V_{(1)})$ with variance
$\mathrm{var}(V_{(1)})$. We thus have
\begin{align}
  \EE(\Delta) &\simeq \int_{-\infty}^{+\infty} |-\EE(V_{(1)})+V_{(1)}|
  \frac{e^{-\frac{(V_{(1)}-\mathsf{E}(V_{(1)}))^2}
      {2\mathrm{var}(V_{(1)})}}}{\sqrt{2\pi \mathrm{var}(V_{(1)})}}
  \D{V_{(1)}} \nonumber \\ &=\sqrt{\frac{2}{\pi} \mathrm{var}(V_{(1)})
  }.
  \label{EDelta.at.Gammac.Grem}
\end{align}
We numerically find $\mathrm{var}(V_{(1)}) \simeq 0.611 J^2$,
therefore, this result is in very good agreement with the value
$\EE(V_{(2)})-\EE(V_{(1)})$ of Eq.~(\ref{gap}), valid in the whole
condensed region.  For the binomial QREM, we can use the probabilities
(\ref{p1.1}) and (\ref{p1.2}) to obtain the exact result
\begin{align}
  \EE(\Delta) &= |-\EE(V_{(1)})-JN| \prob(V_{(1)}=-JN) \nonumber \\
  &\quad + |-\EE(V_{(1)})-J(N-2)| \prob(V_{(1)}=-J(N-2)) \nonumber \\
  &= J\frac{2}{e}.
  \label{EDelta.at.Gammac.Brem}
\end{align}
Note that the distribution functions of $\Delta$ achievable from
Eqs.~(\ref{EDelta.at.Gammac.Grem}) and (\ref{EDelta.at.Gammac.Brem})
are in qualitative agreement, in particular, they confirm the non
self-averaging character of $\Delta$, with the frequencies of the gap
values at $\Gamma=\Gamma_\mathrm{c}$ shown in Figs.~\ref{gb} and
\ref{gg} for finite sizes $N$.

Vice versa, the self-averaging property of the gap in force for
$\Gamma > \Gamma_\mathrm{c}$ implies that, for almost any instance,
$\Delta \simeq \mathsf{E}(\Delta)$ changes smoothly, at finite $N$,
between its value in the condensed region and the normal phase
behavior $\Delta(\Gamma)=2\Gamma$ (exact in the thermodynamic limit or
for $\Gamma \gg \Gamma_\mathrm{c}$), see Figs.~\ref{gb} and \ref{gg}.
According to Eq.~(\ref{E1para}) and in agreement with the numerical
simulations, this change takes place in the interval
$\Gamma_\mathrm{min} \leq \Gamma \leq
\Gamma_\mathrm{min}+2\Gamma_\mathrm{c}/(N-2)$.
Equation~(\ref{Delta(Gamma)}) clearly fails in describing this
behavior.  We have to resort to a more consistent variational analysis
including a superposition of the lowest eigenstates of $\bm{K}$,
namely, $|0_{\bm{K}}\rangle$ and $|1_{\bm{K}}^{(i)}\rangle$,
$i=1,\dots,N$, and of several lowest eigenstates of $\bm{V}$,
$|\bm{n}_l\rangle$ with $l=1,2,\dots$.
Numerical results, not reported here, show that this analysis is able
to capture the quantitative behavior of the gap of each single
instance.  A crucial point, still to be understood, is how the
effective number of lowest $\bm{V}$ eigenstates to be included in this
analysis scales with the size $N$.

\section{Conclusions}
All QREM models are characterized by a universal phase transition
between a condensed phase and a normal one, the critical point
$\Gamma_\mathrm{c}$ being the solution of Eq.~(\ref{critical}).  The
gap averaged over the $\bm{V}$ realizations is finite, and undergoes
the same phase transition, being constant for
$\Gamma<\Gamma_\mathrm{c}$, and growing linearly in $\Gamma$ for
$\Gamma>\Gamma_\mathrm{c}$ (more precisely, as $2\Gamma$, the gap of
the system with $\bm{V}\equiv 0$).  The existence, for each $\bm{V}$
realization, of a minimal gap at $\Gamma=\Gamma_\mathrm{c}$
exponentially small in the system size $N$, implies that, QAd
algorithms (at real or imaginary times) aimed at finding the
configuration of minimal potential, would require a computational time
$t > \Delta_\textrm{min}^{-1}$ exponentially long with $N$ for
\emph{almost any} instance.  In principle, by exploiting the fact that
the average gap is finite, one can hope to elude this limitation by
using an adiabatic path across the $\bm{V}$ ensemble in which the
transit through a gap minimum is avoided~\cite{Farhi:2009}.  However,
it is not clear how to realize this path for $\bm{V}$ of the class
considered here.

Another possibility is the realization of a partial QAd algorithm,
which has the purpose of selecting not the actual lowest eigenstate of
$\bm{V}$ but a set of lowest eigenstates, from which the lowest one
can be found by probabilistic methods. Of course, the success of this
approach depends on the scaling properties of the number of these
eigenstates with respect to the size $N$ of the system.

\section*{Acknowledgement}
M. O. acknowledges CApes for its PNPD program.  The authors declare
that there is no conflict of interest regarding the publication of
this paper.

\section*{References}

\begin{thebibliography}{20}
\expandafter\ifx\csname natexlab\endcsname\relax\def\natexlab#1{#1}\fi
\providecommand{\bibinfo}[2]{#2}
\ifx\xfnm\relax \def\xfnm[#1]{\unskip,\space#1}\fi
\bibitem[{Apolloni et~al.(1989)Apolloni, Carvalho, and
  de~Falco}]{Apolloni:1989}
\bibinfo{author}{B.~Apolloni}, \bibinfo{author}{C.~Carvalho},
  \bibinfo{author}{D.~de~Falco},
\newblock \bibinfo{title}{Quantum stochastic optimization},
\newblock \bibinfo{journal}{Stochastic Processes and their Applications}
  \bibinfo{volume}{33} (\bibinfo{year}{1989}) \bibinfo{pages}{233 -- 244}.
\bibitem[{Finnila et~al.(1994)Finnila, Gomez, Sebenik, Stenson, and
  Doll}]{Finnila:1994}
\bibinfo{author}{A.~B. Finnila}, \bibinfo{author}{M.~A. Gomez},
  \bibinfo{author}{C.~Sebenik}, \bibinfo{author}{C.~Stenson},
  \bibinfo{author}{J.~D. Doll},
\newblock \bibinfo{title}{Quantum annealing: A new method for minimizing
  multidimensional functions},
\newblock \bibinfo{journal}{Chemical Physics Letters} \bibinfo{volume}{219}
  (\bibinfo{year}{1994}) \bibinfo{pages}{343 -- 348}.
\bibitem[{Kadowaki and Nishimori(1998)}]{Kadowaki:1998}
\bibinfo{author}{T.~Kadowaki}, \bibinfo{author}{H.~Nishimori},
\newblock \bibinfo{title}{Quantum annealing in the transverse ising model},
\newblock \bibinfo{journal}{Phys. Rev. E} \bibinfo{volume}{58}
  (\bibinfo{year}{1998}) \bibinfo{pages}{5355--5363}.
\bibitem[{Kirkpatrick et~al.(1983)Kirkpatrick, Gelatt, and
  Vecchi}]{Kirkpatrick:1983}
\bibinfo{author}{S.~Kirkpatrick}, \bibinfo{author}{C.~D.~J. Gelatt},
  \bibinfo{author}{M.~P. Vecchi},
\newblock \bibinfo{title}{{Optimization by Simulated Annealing}},
\newblock \bibinfo{journal}{Science} \bibinfo{volume}{220}
  (\bibinfo{year}{1983}) \bibinfo{pages}{671--680}.
\bibitem[{{Farhi} et~al.(2000){Farhi}, {Goldstone}, {Gutmann}, and
  {Sipser}}]{Farhi:2000}
\bibinfo{author}{E.~{Farhi}}, \bibinfo{author}{J.~{Goldstone}},
  \bibinfo{author}{S.~{Gutmann}}, \bibinfo{author}{M.~{Sipser}},
  \bibinfo{title}{{Quantum Computation by Adiabatic Evolution}},
  \bibinfo{year}{2000}.
\bibitem[{Farhi et~al.(2001)Farhi, Goldstone, Gutmann, Lapan, Lundgren, and
  Preda}]{Farhi:2001}
\bibinfo{author}{E.~Farhi}, \bibinfo{author}{J.~Goldstone},
  \bibinfo{author}{S.~Gutmann}, \bibinfo{author}{J.~Lapan},
  \bibinfo{author}{A.~Lundgren}, \bibinfo{author}{D.~Preda},
\newblock \bibinfo{title}{{A Quantum Adiabatic Evolution Algorithm Applied to
  Random Instances of an NP-Complete Problem}},
\newblock \bibinfo{journal}{Science} \bibinfo{volume}{292}
  (\bibinfo{year}{2001}) \bibinfo{pages}{472--475}.
\bibitem[{Santoro and Tosatti(2006)}]{Santoro:2006}
\bibinfo{author}{G.~E. Santoro}, \bibinfo{author}{E.~Tosatti},
\newblock \bibinfo{title}{Optimization using quantum mechanics: quantum
  annealing through adiabatic evolution},
\newblock \bibinfo{journal}{Journal of Physics A: Mathematical and General}
  \bibinfo{volume}{39} (\bibinfo{year}{2006}) \bibinfo{pages}{R393}.
\bibitem[{J\"org et~al.(2008)J\"org, Krzakala, Kurchan, and Maggs}]{Jorg:2008}
\bibinfo{author}{T.~J\"org}, \bibinfo{author}{F.~Krzakala},
  \bibinfo{author}{J.~Kurchan}, \bibinfo{author}{A.~C. Maggs},
\newblock \bibinfo{title}{Simple glass models and their quantum annealing},
\newblock \bibinfo{journal}{Phys. Rev. Lett.} \bibinfo{volume}{101}
  (\bibinfo{year}{2008}) \bibinfo{pages}{147204}.
\bibitem[{J\"org et~al.(2010)J\"org, Krzakala, Kurchan, Maggs, and
  Pujos}]{Jorg:2010}
\bibinfo{author}{T.~J\"org}, \bibinfo{author}{F.~Krzakala},
  \bibinfo{author}{J.~Kurchan}, \bibinfo{author}{A.~C. Maggs},
  \bibinfo{author}{J.~Pujos},
\newblock \bibinfo{title}{Energy gaps in quantum first-order mean-field–like
  transitions: The problems that quantum annealing cannot solve},
\newblock \bibinfo{journal}{EPL (Europhysics Letters)} \bibinfo{volume}{89}
  (\bibinfo{year}{2010}) \bibinfo{pages}{40004}.
\bibitem[{Adame and Warzel(2015)}]{Warzel2015}
\bibinfo{author}{J.~Adame}, \bibinfo{author}{S.~Warzel},
\newblock \bibinfo{title}{Exponential vanishing of the ground-state gap of the
  quantum random energy model via adiabatic quantum computing},
\newblock \bibinfo{journal}{Journal of Mathematical Physics}
  \bibinfo{volume}{56} (\bibinfo{year}{2015}).
\bibitem[{Battaglia et~al.(2005)Battaglia, Santoro, and
  Tosatti}]{Santoro:2005-2}
\bibinfo{author}{D.~A. Battaglia}, \bibinfo{author}{G.~E. Santoro},
  \bibinfo{author}{E.~Tosatti},
\newblock \bibinfo{title}{Optimization by quantum annealing: Lessons from hard
  satisfiability problems},
\newblock \bibinfo{journal}{Phys. Rev. E} \bibinfo{volume}{71}
  (\bibinfo{year}{2005}) \bibinfo{pages}{066707}.
\bibitem[{Farhi et~al.(2009)Farhi, Goldstone, Gosset, Gutmann, Meyer, and
  Shor}]{Farhi:2009}
\bibinfo{author}{E.~Farhi}, \bibinfo{author}{J.~Goldstone},
  \bibinfo{author}{D.~Gosset}, \bibinfo{author}{S.~Gutmann},
  \bibinfo{author}{H.~B. Meyer}, \bibinfo{author}{P.~Shor}
  (\bibinfo{year}{2009}).
\bibitem[{Ostilli and Presilla(2006)}]{OP:2006}
\bibinfo{author}{M.~Ostilli}, \bibinfo{author}{C.~Presilla},
\newblock \bibinfo{title}{The exact ground state for a class of matrix
  hamiltonian models: quantum phase transition and universality in the
  thermodynamic limit},
\newblock \bibinfo{journal}{Journal of Statistical Mechanics: Theory and
  Experiment} \bibinfo{volume}{2006} (\bibinfo{year}{2006})
  \bibinfo{pages}{P11012}.
\bibitem[{Derrida(1980)}]{Derrida:1980}
\bibinfo{author}{B.~Derrida},
\newblock \bibinfo{title}{Random-energy model: Limit of a family of disordered
  models},
\newblock \bibinfo{journal}{Phys. Rev. Lett.} \bibinfo{volume}{45}
  (\bibinfo{year}{1980}) \bibinfo{pages}{79--82}.
\bibitem[{Johnson et~al.(2005)Johnson, Kemp, and Kotz}]{JKK:2005}
\bibinfo{author}{N.~L. Johnson}, \bibinfo{author}{A.~W. Kemp},
  \bibinfo{author}{S.~Kotz}, \bibinfo{title}{Univariate Discrete
  Distributions}, \bibinfo{publisher}{Wiley}, \bibinfo{edition}{3rd edition}
  edition, \bibinfo{year}{2005}.
\bibitem[{De~Angelis et~al.(1983)De~Angelis, Jona-Lasinio, and
  Sirugue}]{DeAJLS:1983}
\bibinfo{author}{G.~F. De~Angelis}, \bibinfo{author}{G.~Jona-Lasinio},
  \bibinfo{author}{M.~Sirugue},
\newblock \bibinfo{title}{Probabilistic solution of pauli type equations},
\newblock \bibinfo{journal}{Journal of Physics A: Mathematical and General}
  \bibinfo{volume}{16} (\bibinfo{year}{1983}) \bibinfo{pages}{2433}.
\bibitem[{De~Angelis et~al.(1998)De~Angelis, Jona-Lasinio, and
  Sidoravicius}]{DeAJL:1998}
\bibinfo{author}{G.~F. De~Angelis}, \bibinfo{author}{G.~Jona-Lasinio},
  \bibinfo{author}{V.~Sidoravicius},
\newblock \bibinfo{title}{Berezin integrals and poisson processes},
\newblock \bibinfo{journal}{Journal of Physics A: Mathematical and General}
  \bibinfo{volume}{31} (\bibinfo{year}{1998}) \bibinfo{pages}{289}.
\bibitem[{Beccaria et~al.(1999)Beccaria, Presilla, De~Angelis, and
  Jona-Lasinio}]{BPDeAJL:1999}
\bibinfo{author}{M.~Beccaria}, \bibinfo{author}{C.~Presilla},
  \bibinfo{author}{G.~F. De~Angelis}, \bibinfo{author}{G.~Jona-Lasinio},
\newblock \bibinfo{title}{An exact representation of the fermion dynamics in
  terms of poisson processes and its connection with monte carlo algorithms},
\newblock \bibinfo{journal}{EPL (Europhysics Letters)} \bibinfo{volume}{48}
  (\bibinfo{year}{1999}) \bibinfo{pages}{243}.
\bibitem[{Ostilli and Presilla(2017)}]{OP:condensation}
\bibinfo{author}{M.~Ostilli}, \bibinfo{author}{C.~Presilla},
\newblock \bibinfo{title}{First-order quantum phase transitions as
  condensations in the space of states},
\newblock \bibinfo{journal}{arXiv:1712.05294}  (\bibinfo{year}{2017}).

\end{thebibliography}

\end{document}